%% file: paper.tex
\documentclass[twocolumn,superscriptaddress,amsmath,amssymb]{revtex4}
\usepackage[english]{babel}
\usepackage{graphicx}
\usepackage{amssymb}
\usepackage{amsmath}

\begin{document}

\title{Coherent backscattering in the Fock space of ultracold bosonic atoms}

\author{Peter Schlagheck}
\affiliation{D\'epartement de Physique, University of Liege, 4000 Li\`ege, Belgium}
\author{Julien Dujardin}
\affiliation{D\'epartement de Physique, University of Liege, 4000 Li\`ege, Belgium}

\begin{abstract}
We present numerical evidence for the occurrence of coherent backscattering
in the Fock space of a small disordered Bose-Hubbard system consisting of
four sites and containing five particles.
This many-body interference phenomenon can most conveniently be seen in time
evolution processes that start from a Fock state of the Bose-Hubbard system.
It manifests itself in an enhanced detection probability of this initial state
as compared to other Fock states with comparable total energy.
We argue that coherent backscattering in Fock space can be experimentally 
measured with ultracold bosonic atoms in optical lattices using state-of-the-art
single-site detection techniques.
A synthetic gauge field can be induced in order to break time-reversal symmetry
within the lattice and thereby destroy coherent backscattering.
While this many-body interference effect is most prominently visible in the
presence of eigenstate thermalization, we briefly discuss its significance 
in the opposite regime of many-body localization.
\end{abstract}

\maketitle

\section{Introduction}

The eigenstate thermalization hypothesis can be considered to be a
fundamental cornerstone of quantum statistical physics
\cite{Deu91PRA,Sre94PRE,DalO16AdvP}.
It basically assumes for a quantum many-body system that its eigenstates 
are equidistributed in phase space along the energy shell associated 
with the corresponding eigenvalue (provided the total energy is the only 
constant of motion of the system \cite{RigDunOls08N}).
As a consequence, the time average of an observable that is evaluated
along the evolution of a many-body wave packet of this system can be 
determined from the phase-space average of this observable along the 
energy shells that are covered by this wave packet \cite{DalO16AdvP}, 
which is referred to as ergodicity.

While the eigenstate thermalization hyopthesis can safely be taken for 
granted in dilute and weakly interacting three-dimensional quantum gases, 
it is challenged and ultimately invalidated e.g.\ in strongly disordered
one-dimensional many-body systems due to the occurrence of many-body
localization \cite{AltO97PRL,GorMirPol05PRL,BasAleAlt06AP,OgaHus07PRB,ZniProPre08PRB,BarPolMoo12PRL,SerPapAba13PRL,VosAlt13PRL,NanHus15ARCMP,ChoO16S}.
The latter can, in this particular context, be understood as a generalization
of Anderson localization \cite{And58PR} to the many-body domain.
Indeed, all single-particle eigenmodes of an infinite one-dimensional 
disordered system can be shown to be spatially localized even for weak 
disorder \cite{And58PR}, and the presence of a moderate two-body interaction 
may not be sufficient to efficiently couple spatially distant eigenmodes
with each other.
In that case, a many-body wave packet cannot explore the entire energy shell
and will remain localized in the vicinity of the phase space region in which
it was initially prepared.

The present contribution is concerned with a less spectacular and more subtle
localization phenomenon.
It arises in the regime of eigenstate thermalization, where we should expect
ergodicity to occur, and can be seen as the extension of the concept of
\emph{weak localization} to the many-body domain.
From a semiclassical point of view, weak localization is essentially caused
by the constructive interference between backscattered classical paths of the
system and their time-reversed counterparts.
This latter phenomenon of coherent backscattering was experimentally observed
in a number of physical contexts, ranging from laser light scattering off
disordered media \cite{VanLag85PRL,WolMar85PRL} to, more recently, atomic
Bose-Einstein condensates moving in optical speckle fields \cite{JenO12PRL}.
It implies a slight enhancement of the reflection and, in turn, also a slight
reduction of the transmission probability \cite{RicSie02PRL} in disordered
or chaotic wave scattering configurations.

Translated to the many-body domain, coherent back\-scattering gives rise to a 
slight but significant deviation from quantum ergodicity, since the initial 
state of a many-body wave packet thereby has an enhanced probability to be
encountered during the time evolution of this wave packet than any other 
many-body state on the same energy shell.
This many-body coherent backscattering phenomenon was promoted and discussed
in detail in Refs.~\cite{EngO14PRL,EngUrbRic16PTRSA,EngO14TCA,EngO14xxx}
considering both bosonic \cite{EngO14PRL,EngUrbRic16PTRSA} and fermionic
systems \cite{EngO14TCA,EngO14xxx}.
In Ref.~\cite{EngO14PRL}, in particular, it was argued and confirmed through
numerical simulations that a disordered many-body Bose-Hubbard system which
is initially prepared in one of its Fock states will, in the course of time 
evolution, be encountered and detected about twice as often in this initial 
state as compared to any other Fock state with comparable total energy.

The main purpose of this paper is to investigate this delicate but robust 
many-body interference phenomenon within a relatively small system that can 
possibly be realized in an experiment using ultracold bosonic quantum gases.
We specifically consider for this purpose a disordered Bose-Hubbard ring 
containing four sites, which describes a single plaquette of a two-dimensional 
optical square lattice exhibiting random on-site energies.
This Bose-Hubbard ring is supposed to be loaded with five particles.
As we shall show below, the effect of coherent backscattering in Fock space 
can indeed be encountered in such a microscopic many-body system, and its 
observation with ultracold bosonic atoms is within the reach of present-day 
state-of-the-art experiments.

A proposal for the experimental protocol that can be implemented for this
purpose is outlined and discussed in Section~\ref{sec:exp}.
Section~\ref{sec:cbs} discusses the evolution of coherent backscattering 
in Fock space as a function of time as well as its dephasing in the presence
of an artificial gauge field that induces complex inter-site hopping matrix
elements within the Bose-Hubbarg ring and thereby breaks time-reversal 
invariance.
In Section~\ref{sec:cl} we describe how to define and numerically compute
classical predictions for the average detection probabilities of Fock
states, using the van Vleck--Gutzwiller propagator and the diagonal 
approximation.
Finally, Section~\ref{sec:mbl} discusses to which extent coherent 
backscattering in Fock space plays a role also in the regime of many-body 
localization.

\section{The proposed experimental procedure}

\label{sec:exp}

We consider a finite number of ultracold bosonic atoms that are confined 
within a single plaquette of a two-dimensional optical lattice.
This plaquette can be isolated from the rest of the lattice by means of
a red-detuned laser beam that is perpendicularly focused on the lattice
or by generating a two-dimensional superlattice, in close analogy with 
Refs.~\cite{FoeO07N,CheO08PRL}.
This latter approach would yield a periodic checkerboard 
arrangement of plaquettes within which near-identical experiments can be
performed in parallel.

Within the single-band approximation, the many-body dynamics of this 
configuration can be described by the Bose-Hubbard Hamiltonian
\begin{equation}
  \hat{H} = \sum_{l=1}^L \left[ \epsilon_l \hat{b}_l^\dagger \hat{b}_l 
  - J \left( e^{i\varphi} \hat{b}_l^\dagger \hat{b}_{l-1} +
  e^{-i\varphi} \hat{b}_{l-1}^\dagger \hat{b}_{l} \right)
  + \frac{U}{2} \hat{b}_l^\dagger\hat{b}_l^\dagger \hat{b}_l\hat{b}_l\right]
  \label{eq:H}
\end{equation}
with $L=4$, where $\hat{b}_l^\dagger$ and $\hat{b}_l$ respectively represent the
bosonic creation and annihilation operators associated with the site $l$ 
(we formally identify $\hat{b}_0\equiv\hat{b}_L$, 
$\hat{b}_0^\dagger\equiv\hat{b}_L^\dagger$).
$\epsilon_l$ denotes the on-site energy of the site $l=1,\ldots,L$,
$U$ represents the on-site interaction energy between two atoms, and 
$J$ is the hopping amplitude between adjacent sites.
Time-reversal symmetry can be broken by imposing a finite hopping phase
$\varphi\neq 0$ which can be engineered through the application of a 
synthetic gauge field \cite{LinO09N,GolO10PRL,DalO11RMP}.

The plaquette is to be loaded with a finite number of atoms in the
deep Mott insulator regime where inter-site hopping can be neglected.
This loading process is supposed to take place in such a way that a 
well-defined single Fock state in the site basis is reliably obtained 
as initial state of this many-body system for a suitable choice of the 
on-site energies $\epsilon_l$ \cite{rem_initial}.
For the sake of definiteness, we consider the initial state 
$|\mathbf{n}^{(i)}\rangle\equiv|n_1^{(i)},n_2^{(i)},n_3^{(i)},n_4^{(i)}\rangle 
= |1,2,2,0\rangle$ in the following, where $n_l$ denotes the number of 
atoms located on the site $l$.

At time $t=0$ inter-site hopping is switched on, such that the atoms can
move from site to site.
We furthermore suppose that the on-site energies of the individual sites
acquire random values at $t=0$, which in practice can be achieved by means
of an optical speckle field \cite{LyeO05PRL,BilO08N}.
For the sake of numerical convenience, we assume for the following that
for $t>0$ the on-site energies are given by uncorrelated random values 
$\epsilon_l \in[0,W]$ that result from a uniform probability distribution
characterized by the energy scale $W$ \cite{rem_disorder}.

The inter-site hopping is to be switched off at a given final time $T>0$, 
such that the sites are again isolated.
Subsequently, the populations of the individual sites of the lattice have to
be detected site by site, yielding a result that can be represented by 
the final Fock state $|\mathbf{n}\rangle\equiv|n_1,n_2,n_3,n_4\rangle$.
From the experimental point of view, this latter task appears to represent 
the most challenging part of this proposal.
It could possibly be achieved by combining the high-resolution imaging 
techniques pioneered in Refs.~\cite{BakO09N,She10N}, which allow to detect 
binary populations (i.e., $n_l=0$ or $1$) of individual lattice sites,
with a free expansion of the atoms onto previously unpopulated sites of the
two-dimensional optical lattice, in close analogy with Ref.~\cite{KauO16S}.

The idea is now to repeat this particular experiment many times, each time
starting with the preparation of the same initial state 
$|\mathbf{n}^{(i)}\rangle$, in order to measure the probabilities for 
obtaining a given final state $|\mathbf{n}\rangle$ with sufficient 
statistical accuracy.
A disorder average is to be introduced by randomly varying the on-site 
energies $\epsilon_l$ from one experiment to the next one.

\begin{figure}
  \includegraphics[width=\columnwidth]{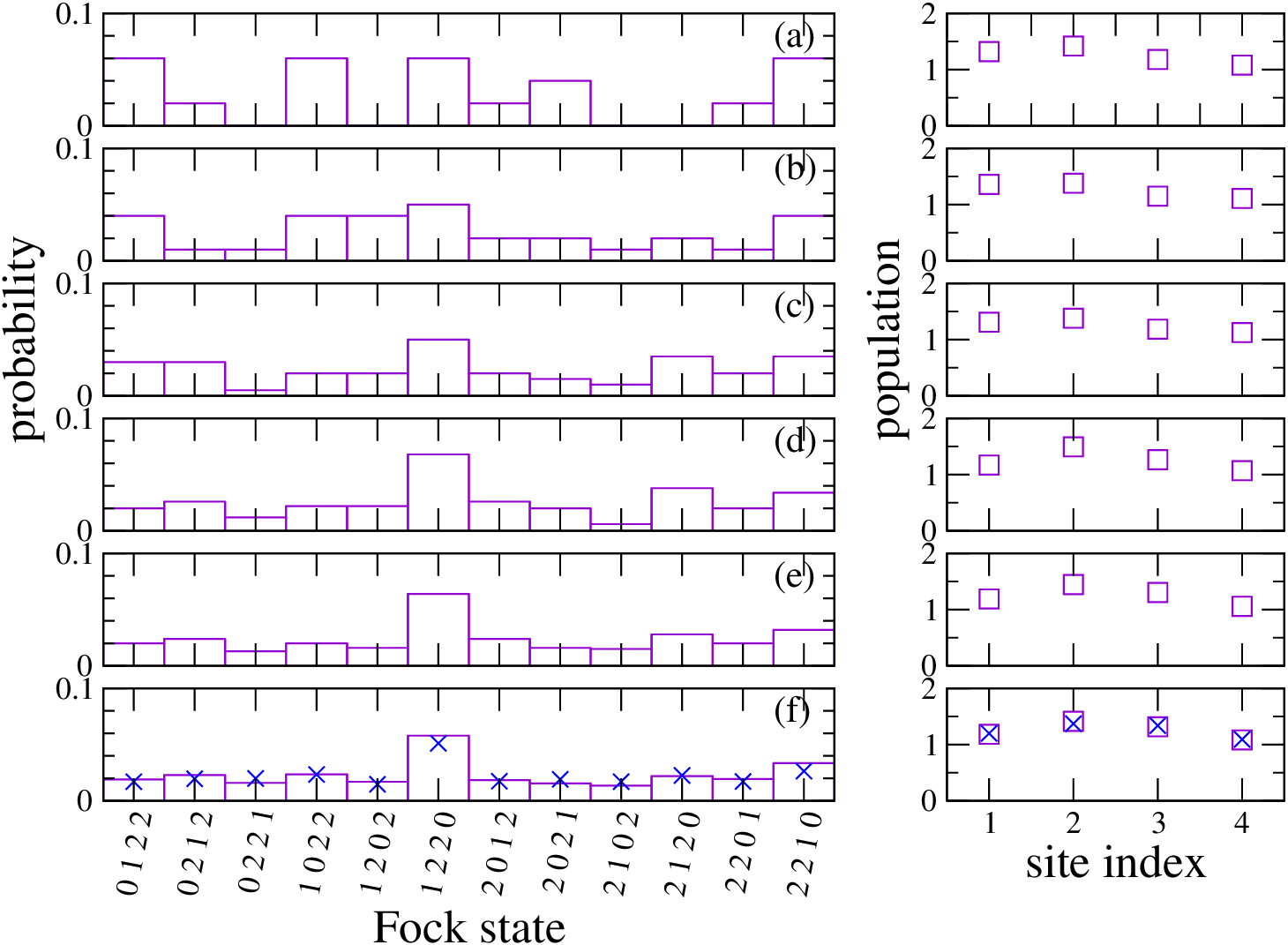}%
  \caption{\label{fig:meas}
    Numerical simulation of a measurement process.
    Plotted are, in the left column, the probabilities for detecting a 
    given Fock state and, in the right column, the average populations
    on each site of the lattice.
    These data are determined from the accumulated measurement outcomes 
    that are obtained after (a) 50, (b) 100 (c) 200, (d) 500, (e) 1000, 
    and (f) 2000 repetitions of the numerical experiment.
    The blue crosses in (f) show the average quantum probabilities (left) and
    populations (right) obtained from the numerical integrations of the time 
    evolution generated by the Bose-Hubbard Hamiltonian \eqref{eq:H}.
    The calculation was done for the initial state $|1,2,2,0\rangle$,
    the lattice parameters $U=J$, $\varphi=0$, $W=4J$, and the evolution
    time $T = 10 \hbar/J$.
    Only Fock states that have the same interaction energy as the initial
    state are shown in the left column.
  }
\end{figure}

Figure~\ref{fig:meas} shows a numerical simulation of the outcome resulting 
from such a sequence of experiments, for the parameters $U=J$, $W=4J$, 
$\varphi=0$, and for the final time $T = 10 \hbar/J$.
This simulation is performed by means of numerical integrations of the time 
evolution generated by the Bose-Hubbard Hamiltonian \eqref{eq:H} 
\cite{rem_decay}, each one being carried out for a given random choice of 
on-site energies.
The resulting quantum probabilities $p_{\mathbf{n}}$ for obtaining a particular
final state $|\mathbf{n}\rangle$ are then used to simulate a measurement
process corresponding to a projection onto one of those Fock states.

We see that some 1000 repetitions of this experiment would be necessary
in order to achieve statistical convergence.
At that point, all Fock states $|\mathbf{n}\rangle\neq|\mathbf{n}^{(i)}\rangle$ 
that exhibit the same interaction energy as the initial state 
$|\mathbf{n}^{(i)}\rangle = |1,2,2,0\rangle$ are detected with about equal 
probability, which indicates that we are in a regime of eigenstate 
thermalization.
There is one exception, however:
the initial state $|\mathbf{n}^{(i)}\rangle$ is significantly more often 
detected than those other Fock states.

This notable deviation from quantum ergodicity in Fock space seems to have 
a tiny but significant impact onto the average populations per site, as can 
be seen in the right column of Fig.~\ref{fig:meas}.
While the naive application of the ergodic hypothesis would straightforwardly 
yield a uniform average distribution of atoms within the plaquette, 
the populations on the sites $l=2$ and $3$ are found to be enhanced 
by about 20 percent as compared to the sites $l=4$ and $l=1$, respectively, 
which is consistent with the initial state $|1,2,2,0\rangle$.
Obviously, much less repetitions of the experiment are required in order to
obtain a satisfactory statistical convergence of the average populations per 
site as compared to the probabilities in Fock space.
We note that those average populations can be measured with a simpler 
experimental scheme than the one that we outline above, namely by performing 
many near-identical experiments in parallel within a two-dimensional 
superlattice configuration, and by generalizing the Brillouin zone mapping 
technique used in Refs.~\cite{FoeO07N,CheO08PRL} to two spatial dimensions, 
such that the populations of the sites can be clearly distinguished in 
momentum space after a free expansion of the atomic cloud.

\section{Coherent backscattering in Fock space}

\label{sec:cbs}

\begin{figure}
  \includegraphics[width=\columnwidth]{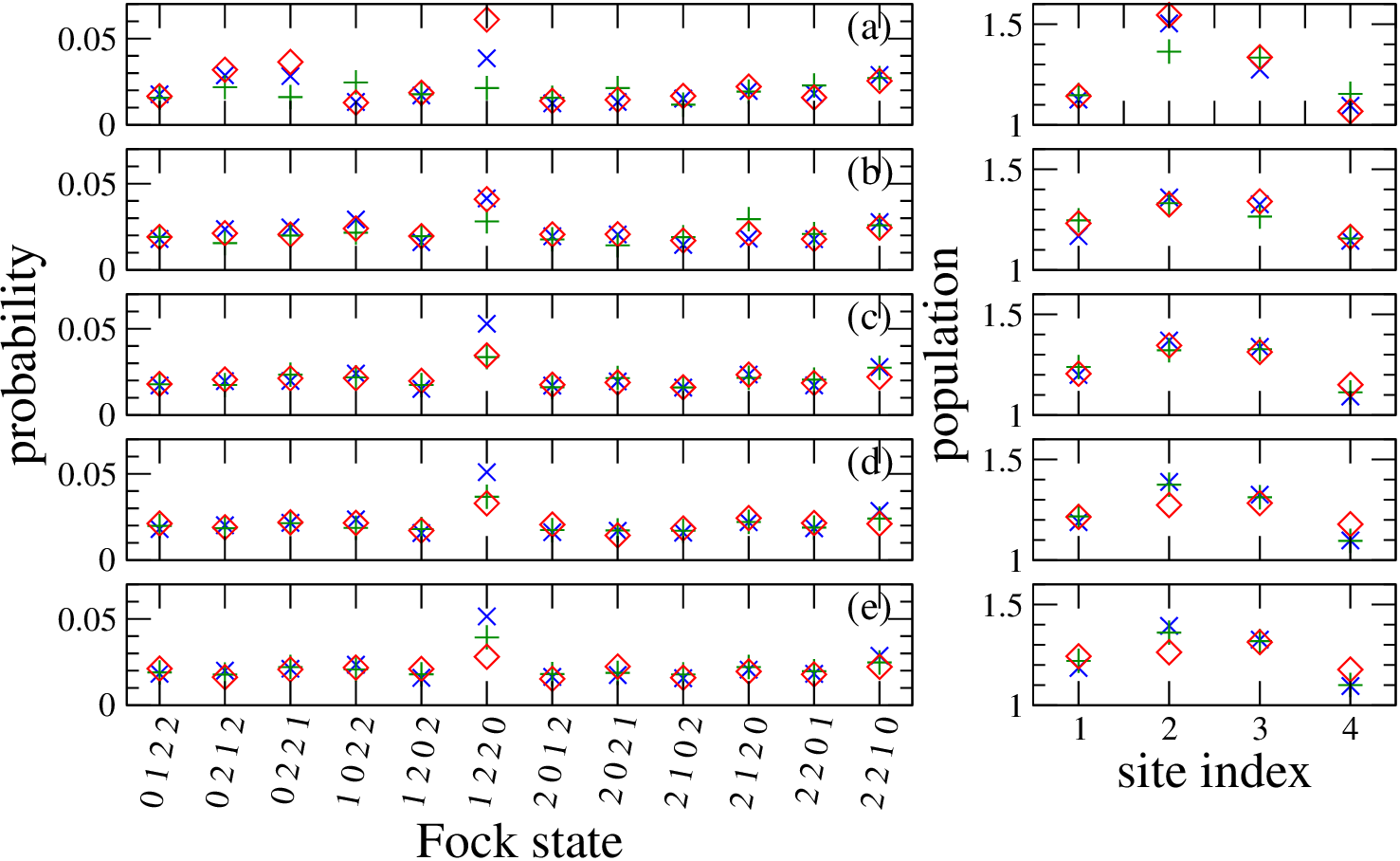}%
  \caption{\label{fig:evol}
    Time evolution of the probabilities for detecting a given Fock state 
    (left column) and of the average populations per site (right column),
    shown for $J t / \hbar = 2$ (a), $5$ (b), $10$ (c), $20$ (d), and $50$ (e).
    The quantum calculations were done for the initial state $|1,2,2,0\rangle$,
    the lattice parameters $U=J$, $W=4J$, and the hopping phases $\varphi=0$
    (blue $\times$) as well as $\varphi=0.1 \pi$ (green $+$).
    Red diamonds show the corresponding classical predictions for $\varphi=0$.
  }
\end{figure}

To confirm that this signature of non-ergodicity is not a transient
phenomenon that would disappear after longer evolution times, we show in
Fig.~\ref{fig:evol} snapshots of the time evolution of the probabilities 
for detecting a given Fock state, for the same initial state $|1,2,2,0\rangle$
and the same lattice parameters $U=J$, $W=4J$ as in Fig.~\ref{fig:meas}.
We clearly see that after an initial transient regime, which is very specific
to the initial state under consideration and which lasts about one 
hopping period $\tau=2\pi/J$, a stable and practically stationary probability 
distribution is obtained in the Fock space. 
As in Fig.~\ref{fig:evol}, the initial state is characterized by a significant
enhancement of its detection probability as compared to other Fock states
with comparable interaction energy, the latter being detected with equal
probability.

The origin of this enhancement can, to a dominant extent, be attributed to
coherent backscattering in Fock space.
This latter phenomenon generally refers to an enhanced probability for
backscattering of a wave that propagates in the presence of chaos or disorder,
owing to the systematic and robust constructive interference between 
backscattered paths and their time-reversed counterparts \cite{AkkMon}.
It is at the origin of weak localization in mesoscopic electronic conduction
process \cite{AkkMon} and was recently observed with ultracold bosonic matter
waves \cite{JenO12PRL,LabO12EPL}.
As was explained in Ref.~\cite{EngO14PRL}, the generalization of the concept
of coherent backscattering to a many-body Bose-Hubbard lattice can be done
by considering each site of the lattice as being an independent oscillator-like
degree of freedom.
Classical ``paths'' within such a lattice correspond then to solutions of
a discrete Gross-Pitaevskii equation, as we shall further elaborate in the 
subsequent section.
Backscattered paths start out with a given set of amplitudes per site (namely
the ones that correspond to the initial state $|\mathbf{n}^{(i)}\rangle$)
and end up with exactly the same amplitudes, while associated phases may
be arbitrary.

A characteristic signature of coherent backscattering and weak localization
is its sensitivity to any breaking of time-reversal invariance induced by 
means of a weak gauge field (which in the electronic case would most naturally 
be given by a magnetic field).
While such a gauge field may not necessarily modifiy the classical paths in
an appreciable manner, it does affect the associated phases in such a way 
that the interference between a backscattered path and its time-reversed 
counterpart is no longer entirely constructive.
As a consequence, the coherent backscattering amplitude generally decreases 
with increasing strength of the gauge field.

\begin{figure}
  \includegraphics[width=\columnwidth]{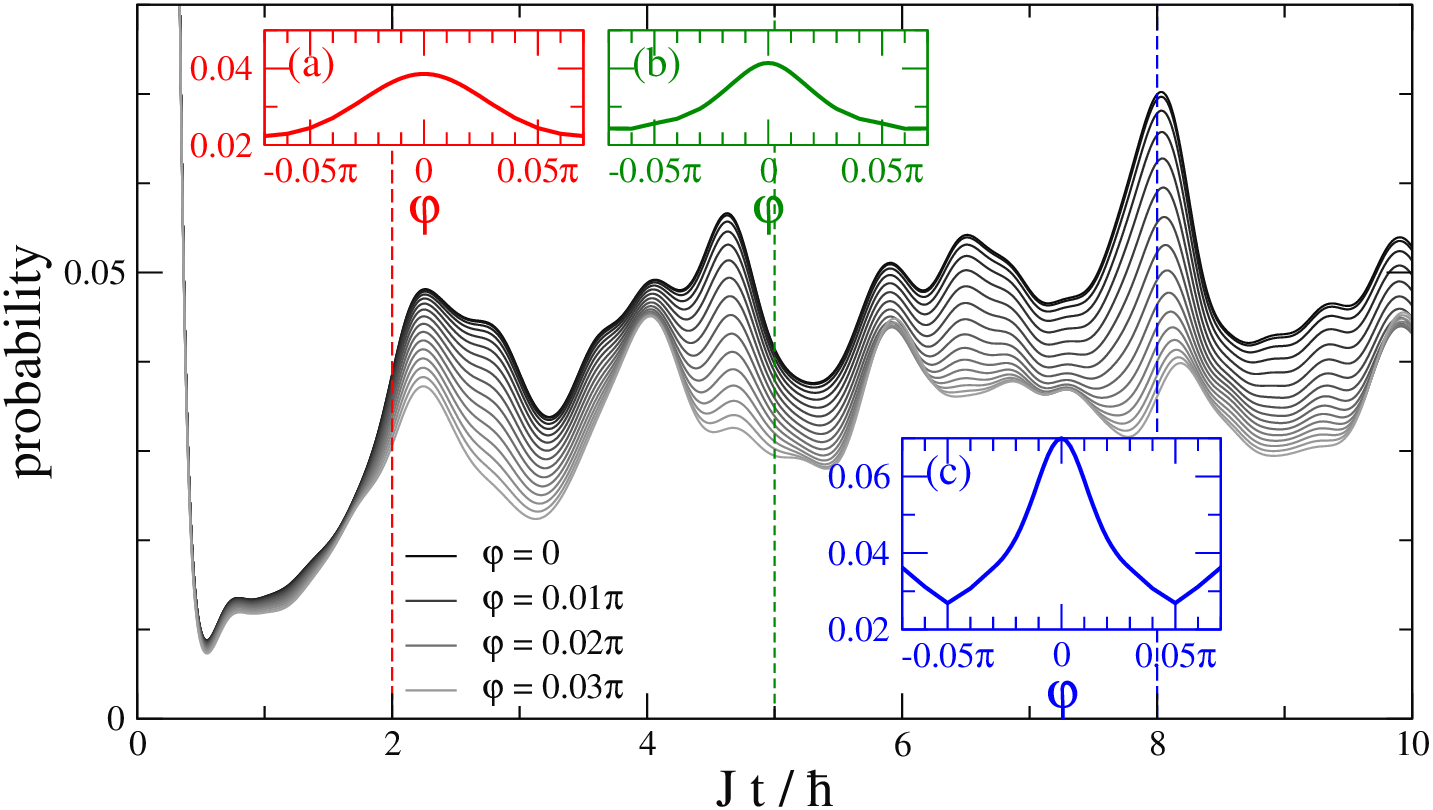}%
  \caption{\label{fig:deph}
    Time evolution of the probability to detect the initial state
    $|1,2,2,0\rangle$ plotted for the hopping phases $\varphi = 0$ (darkest
    curve), $0.002$, $0.004$, \ldots $0.03$ (lightest curve) at the lattice 
    parameters $U=J$, $W=4J$.
    The insets show these probabilities as a function of the hopping phase
    $\varphi$ at fixed times $t = 2 \hbar/J$ (left inset, shown in red),
    $t = 5 \hbar/J$ (middle inset, shown in green), and $t = 8 \hbar/J$ 
    (right inset, shown in blue).
  }
\end{figure}

This dephasing behaviour is indeed found in our four-sites Bose-Hubbard ring.
As can be seen in Fig.~\ref{fig:evol}, the presence of a weak hopping phase
$\varphi=0.1\pi$ significantly reduces the probability for detecting the
initial state as compared to the time-reversal invariant case $\varphi=0$,
whereas the probabilities for detecting other Fock states 
$|\mathbf{n}\rangle\neq|\mathbf{n}^{(i)}\rangle$ are not appreciable modified.
Figure~\ref{fig:deph} shows the probability to detect the initial state
as a function of time for various hopping phases.
We see in the insets that the backscattering probability continuously 
decreases with increasing hopping phase for $|\varphi| \lesssim 0.05\pi$, 
which is a clear signature of weak localization.

\section{Comparison with classical predictions}

\label{sec:cl}

To ultimately confirm that the enhancement of the detection probability
of the initial state as compared to other Fock states with comparable
total energy arises due to a quantum interference phenomenon taking place 
in the Fock space of the bosonic many-body system, we compare the numerically
computed detection probabilities for the final state 
$|\mathbf{n}\rangle\equiv|n_1,\ldots,n_L\rangle$ with classical predictions 
that precisely neglect the presence of those many-body interferences.
To this end, we employ the framework of the semiclassical van Vleck--Gutzwiller
propagator \cite{Gut}.
The latter allows one to express the matrix elements of the time evolution 
operator of the bosonic many-body system in its Fock basis as a sum over 
classical trajectories that evolve according to the discrete nonlinear 
Schr\"odinger equation
\begin{eqnarray}
  i \hbar \frac{d}{dt} \psi_l(t) & = & \epsilon_l \psi_l(t)
  - J \left[\psi_{l+1}(t) + \psi_{l-1}(t)\right] \nonumber\\&&
  + U \left[ |\psi_l(t)|^2 - 1 \right] \psi_l(t) \label{eq:gp}
\end{eqnarray}
(with $\psi_0 \equiv \psi_L$) 
and exhibit the initial and final conditions
$|\psi_l(0)|^2 = n_l^{(i)}+1/2$ and $|\psi_l(T)|^2 = n_l+1/2$
for all $l=1,\ldots,L$.
Each one of those trajectories contributes to the propagator with a complex 
term whose amplitude is related to the classical stability of the trajectory
(i.e., to the Lyapunov exponent in case of chaotic motion) and whose phase
is given by the classical action of the trajectory.

Detection probabilities in Fock space are obtained from the square moduli
of those matrix elements of the van Vleck--Gutzwiller propagator.
The resulting double sum over the above trajectories is substantially
simplified upon disorder average, as ``off-diagonal'' terms in this double 
sum, involving the contributions of interferences between two different 
trajectories, exhibit complex phase factors that rapidly vary under a 
variation of the on-site energies $\epsilon_l$ and are therefore
averaged to zero as soon as a disorder average is performed.
In leading order, we are then left with the \textit{diagonal approximation}
which only accounts for pairings of trajectories with themselves in the
above double sum.
As was elaborated in Ref.~\cite{EngO14PRL}, the detection probability for
the final Fock state $|\mathbf{n}\rangle$ can then be expressed as
\begin{equation}
  p_\mathbf{n}^{(\rm diag)} = \sum_\gamma \left|  \det 
  \frac{\partial (\theta_2^{(i)},\ldots,\theta_L^{(i)})_\gamma}
  {\partial (I_2^{(f)},\ldots,I_L^{(f)})}\right| \,,
  \label{eq:psc}
\end{equation}
where we introduce according to 
$\psi_l(0) \equiv [I_l^{(i)}]^{1/2} \exp[ i\theta_l^{(i)}]$ and 
$\psi_l(T) \equiv [I_l^{(f)}]^{1/2} \exp[i\theta_l^{(f)}]$ the
action-angle variables $(I_l^{(i)},\theta_l^{(i)})$ and $(I_l^{(f)},\theta_l^{(f)})$
that are associated with the initial and final classical amplitudes 
on site $l$, respectively.
The Jacobi determinant arising in Eq.~\eqref{eq:psc} excludes one of the
sites (which we choose to be site $1$ without loss of generality) in order
to account for the presence of a cyclic phase that is associated with the
total number of particles being a constant of motion \cite{EngO14PRL}.

In the presence of fully developed chaos and ergodicity, we can reformulate
Eq.~\eqref{eq:psc} as
\begin{equation}
  p_{\mathbf{n}}^{(\rm diag)} = \int_0^{2\pi} \frac{d\theta_2}{2\pi} \cdots 
  \int_0^{2\pi} \frac{d\theta_L}{2\pi} \prod_{l=2}^L 
  \delta\left[ I_l^{(f)} - |\psi_l(T)|^2\right] \label{eq:pch}
\end{equation}
using the Hannay--Ozorio de Almeida sum rule \cite{HanOzo84JPA}.
Here $\psi_l(t)$ describes the complex classical amplitude on site $l$ 
which evolves according to Eq.~\eqref{eq:gp} and emanates from the initial
conditions $\psi_l(0) = [n_l^{(i)}+1/2]^{1/2} \exp[ i\theta_l^{(i)}]$ where we 
set $\theta_1^{(i)} = 0$ without loss of generality.
Employing the box regularization $\delta(x)= \theta(d/2-|x|)/d$
of the delta function with the width $0 < d \ll 1$ (where $\theta$ denotes 
the Heavyside step function),
we can numerically solve Eq.~\eqref{eq:pch} by means of a Monte-Carlo method.
$p_{\mathbf{n}}^{(\rm diag)}$ is then obtained in practice by counting the number 
of Monte-Carlo trajectories that match the required final conditions 
$|\psi_l(T)|^2 = n_l+1/2$ for $l=2,\ldots,L$ with the tolerance $\pm d/2$
and by dividing this number by $d^{L-1}$ and by the total number of trial
trajectories.

However, care has to be taken as the classical dynamics of
the Bose-Hubbard system may not always be fully ergodic.
Nonergodicity typically arises if for some disorder realization the on-site 
energy on one of the sites is substantially higher or lower than the on-site 
energies of the other sites, in which case this particular site can be 
considered as being essentially disconnected from the rest of the lattice.
The associated angle variable then represents another cyclic variable, which
implies that this site has to be excluded in the integrals and the
product on the right-hand side of Eq.~\eqref{eq:pch}.
As a consequence, the associated contribution to $p_{\mathbf{n}}^{(\rm diag)}$
involves a division by $d^{L-2}$ and not $d^{L-1}$.
Evidently, this reasoning straightforwardly generalizes for the case that
two or more sites are energetically disconnected from the lattice.

In order to properly account for such nonergodic situations, we numerically
determine the effective dimension of the Lagrangian manifold for each
trajectory that properly matches the required final conditions.
We use the Grassberger-Procaccia algorithm for this purpose 
\cite{GraPro83PRL,GraPro83PsD}, i.e., we generate a long time series
emanating from the corresponding initial conditions (involving $1000$ time 
steps with the temporal spacing $\Delta t = 0.5 \hbar / J$) and determine 
the correlation dimension of the resulting set of phase-space points.
This dimension is then compared with the correlation dimensions that are 
obtained for a set of randomly chosen phase-space points 
$(\psi_1,\ldots,\psi_L)$ where we impose either no restrictions apart from
the conservation of the total number of particles (in which case the resulting
set mimicks perfect ergodicity) or additional restrictions concerning the
amplitudes on one or several sites.
The result of this comparison determines then the power of the tolerance
parameter $d$ that we have to use in order to properly count the contribution
of the trajectory under consideration.

The resulting classical Fock state probabilities obtained for $d=0.05$ 
are displayed as red diamonds in Fig.~\ref{fig:evol}.
After the initial transient regime, they agree very well with the quantum 
probabilities for the states $|\mathbf{n}\rangle\neq|\mathbf{n}^{(i)}\rangle$
that have the same interaction energy as the initial state.
The classical diagonal approximation still predicts a slight enhancement
of the probability for detecting the initial state $|\mathbf{n}^{(i)}\rangle$
as compared to those other Fock states.
This enhancement is, however, much less pronounced than the one that is found
for the exact quantum probability $p_{\mathbf{n}^{(i)}}$ (in the absence of time 
reversal symmetry breaking), the latter being about twice as large as
the classical prediction $p_{\mathbf{n}^{(i)}}^{(\rm diag)}$ for long evolution
times $T \gg \hbar / J$.
This confirms that coherent backscattering, which represents the leading 
correction to the diagonal approximation [Eq.~\eqref{eq:psc}] for the
case $\mathbf{n}=\mathbf{n}^{(i)}$, is at the origin of the enhanced detection
probability of the initial state \cite{rem_retraced}.

\section{Interplay with many-body localization}

\label{sec:mbl}

The right column of Fig.~\ref{fig:evol} shows that the occurrence of 
coherent backscattering in Fock space has a nonvanishing but quantitatively 
rather tiny impact on the average populations on the lattice sites, which
appears to be too small to be detected in an experiment.
In particular, the inhomogeneous distribution of the populations, which 
reflects the ``memory'' of the initial state in the time evolution, cannot
be entirely traced back to this particular quantum many-body interference 
phenomenon.
Instead, as we can judge from the classical population distributions in the 
right column of Fig.~\ref{fig:evol}, it seems to dominantly arise from a 
limitation of the effectiveness of classical ergodicity, due to the fact that 
for a significant fraction of disorder realizations (and initial on-site angle 
variables, as discussed in the previous section) the classical energy shell 
associated with the initial Fock state cannot be efficiently explored and 
covered within the time scale under consideration.
Accordingly, we find in the left column of Fig.~\ref{fig:evol} that also 
the classical prediction for the detection probability of the initial 
state $|1,2,2,0\rangle$ is slightly enhanced with respect to other Fock states
on the same energy shell.
In contrast to the true quantum detection probability of the initial state,
this latter ``classical'' enhancement appears to decrease with increasing
evolution time as we see in Fig.~\ref{fig:evol}(e), in accordance with the
expectation that a full classical exploration of the energy shell should, on
average, become more and more feasible with increasing evolution time.

\begin{figure}
  \includegraphics[width=\columnwidth]{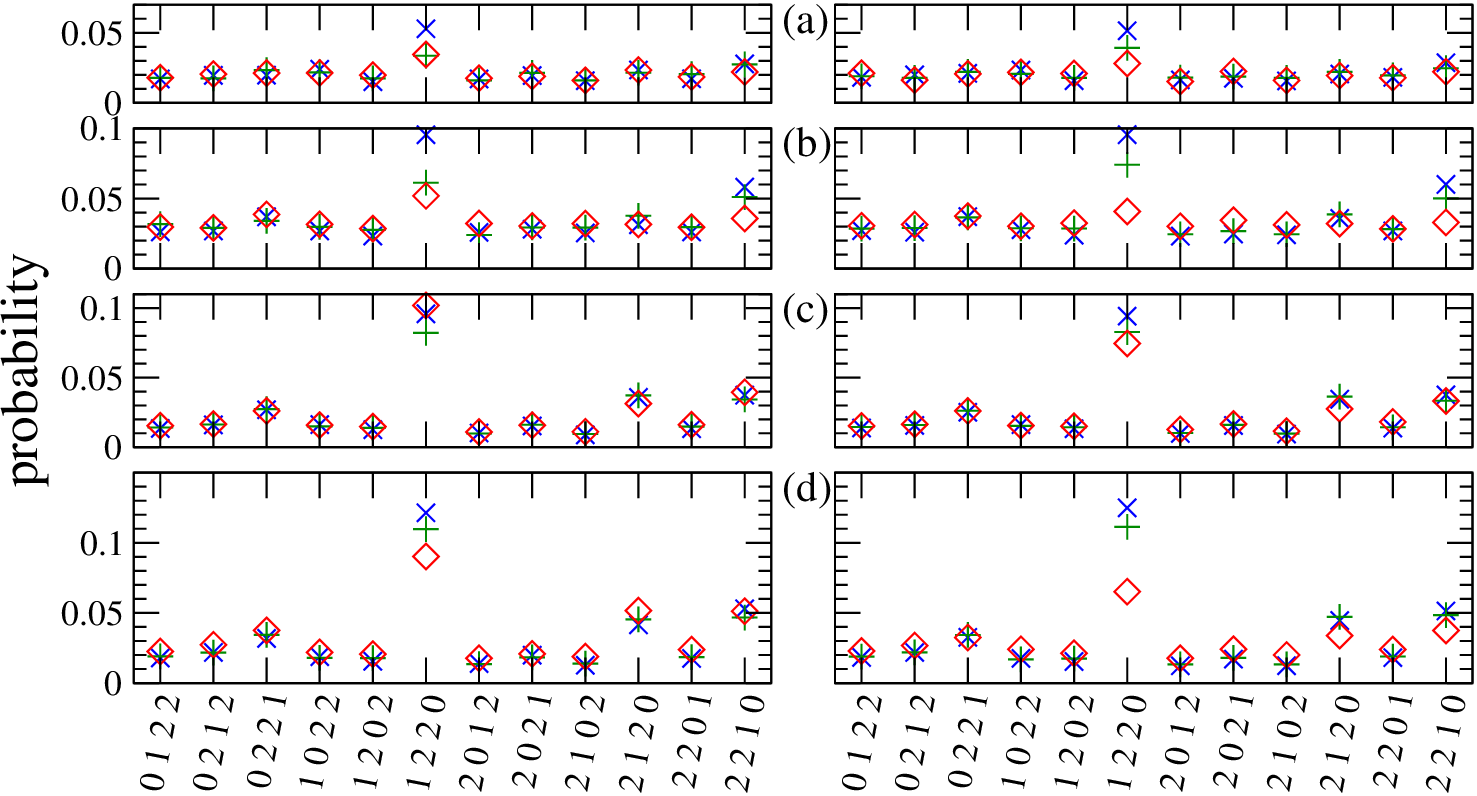}%
  \caption{\label{fig:loc}
    Probabilities for detecting a given Fock state at $J T / \hbar = 10$ 
    (left column) and $J T / \hbar = 100$ (right column) for the lattice 
    parameters (a) $U=J$, $W=4J$, (b) $U=4J$, $W=4J$, (c) $U=J$, $W=10J$, 
    (d) $U=4J$, $W=10J$.
    The quantum calculations were done for the initial state $|1,2,2,0\rangle$
    and the hopping phases $\varphi=0$ (blue $\times$) as well as 
    $\varphi=0.1 \pi$ (green $+$).
    Red diamonds show the corresponding classical predictions for $\varphi=0$.
  }
\end{figure}

As is shown in Fig.~\ref{fig:loc}, this latter tendency to recover classical 
ergodicity at long evolution times becomes more pronounced for larger 
interaction strengths ($U=4J$), due to the thereby enhanced nonlinear 
coupling and mixing of single-particle eigenmodes of the disordered system 
in the discrete Gross-Pitaevskii equation \eqref{eq:gp}.
Furthermore, it plays a prominent role in the presence of stronger disorder
($W=10J$) for which eigenstate thermalization can no longer be taken for 
granted.
In such a situation, we find a highly pronounced enhancement of the 
detection probability of the initial state which exceeds the probabilities
of other states on the same energy shell by about a factor 4 or 5.
However, as we can clearly see in the lower two panels on the left column
of Fig.~\ref{fig:loc}, this latter enhancement dominantly arises, for finite 
evolution times $t \sim 10 \hbar / J$, due to a deficient ergodicity on the
classical level or, more precisely, due to the incapacity of classical
trajectories to explore the entire energy shell within those finite time scales.
The effectively accessible phase space is therefore rather restricted, which
in turn implies, as was pointed out in the supplementary material of 
Ref.~\cite{EngO14PRL}, that the set of trajectories that contribute to the 
backscattering probability mostly contain self-retraced trajectories
which are identical to their time-reversed counterparts.
The quantum return probability to the initial Fock state is, at 
$t \sim 10 \hbar / J$, therefore only very weakly enhanced (if at all) 
as compared to its classical prediction, and it is hardly affected by the 
breaking of time-reversal invariance due to the application of a gauge field.

The right column of Figs.~\ref{fig:loc}(c) and \ref{fig:loc}(d) confirms
that we are concerned here with a typical scenario of many-body localization.
While the classical prediction for the detection probability of the initial 
state decreases at very large evolution times $T\sim 100 \hbar/J$, due to the 
(sub-diffusive) increase of the average classical phase space that can 
effectively be explored within an increasing evolution time 
\cite{She93PRL,PikShe08PRL,FlaKriSko09PRL,KopO08PRL}, the 
quantum probabilities remain essentially frozen at such large times.
We should note that $T = 100 \hbar/J$ exceeds the Heisenberg time 
$\tau \sim \hbar / \Delta$ of the quantum many-body system under consideration,
which for $U=J$ and $W=4J$ exhibits a mean energy spacing of the order of 
$\Delta \sim 0.05 J$ in the central part of its spectrum.
This limits the semiclassical relevance of slow classical (sub-)diffusion 
processes that evolve over such large time scales.

\section{Conclusion}

In summary, we provided numerical evidence that coherent backscattering in 
Fock space can arise within a relatively small many-body system, namely a 
Bose-Hubbard ring that consists of four sites and contains five particles.
The latter can be realized within a single plaquette of a two-dimensional 
optical square lattice which is to be loaded with five bosonic atoms.
The experimental protocol to measure the crucial signature of many-body
coherent backscattering, namely the enhancement of the detection probability
of the initial state as compared to other Fock states with comparable total 
energy, involves as major challenges the implementation of state-of-the-art 
single-site detection techniques \cite{KauO16S} as well as a large number 
of repetitions of the experiment combined with the use of on-site disorder.
In addition, a weak artificial gauge field, corresponding to an inter-site 
hopping that can be as small as $\varphi \simeq \pm 0.02\pi$ (see 
Fig.~\ref{fig:deph}), ought to be induced within the plaquette in order to 
break time-reversal invariance and thereby destroy many-body coherent 
backscattering.

Our study indicates that coherent backscattering in Fock space is most 
significant and effective in a parameter regime where the many-body system
under consideration exhibits eigenstate thermalization.
The interplay with many-body localization was briefly discussed but certainly
deserves further explorations.
In particular, we could speculate that coherent backscattering in Fock space 
might act as some sort of precursor to many-body localization, in a similar 
manner as weak localization being a precursor to strong Anderson localization 
within the single-particle context.
More research in this direction, possibly involving a generalization of
the self-consistent theory of localization \cite{VolWoe82PRL} to the 
many-body domain, appears to be useful in order to clarify this issue and
thereby contribute to our understanding of the transition to many-body 
localization.

\begin{acknowledgements}

We thank Thomas Engl, Klaus Richter, Philipp Preiss, and Juan Diego Urbina 
for useful discussions.

\end{acknowledgements}

\input paper.bbl


\end{document}

%% file: paper.bbl
\providecommand{\WileyBibTextsc}{}
\let\textsc\WileyBibTextsc
\providecommand{\othercit}{}
\providecommand{\jr}[1]{#1}
\providecommand{\etal}{~et~al.}

%% file: paper.bbl
\begin{thebibliography}{[10]}
\bibitem{rem_initial}
Detections of the atomic populations on the lattice sites can be carried out
  right after this preparation process to verify its reliability in practice.
\bibitem{rem_disorder}
We thereby neglect spatial correlations that would necessarily arise within
  disordered potentials realized by speckle fields.
\bibitem{rem_decay}
We do not account for three-body losses that could arise if three atoms are
  located on the same site.


\othercit
\end{thebibliography}
